\documentclass[runinaddress,showpacs,aps,jpcm, twocolumn,superscriptaddress,a4paper,floatfix]{revtex4-1}

\usepackage{amsfonts}
\usepackage{amsmath}
\usepackage{amssymb}
\usepackage{bm}
\usepackage{color}
\usepackage{color,graphicx}
\usepackage{dcolumn}	
\usepackage{gensymb}
\usepackage{graphicx}	
\usepackage[hyperfootnotes,hyperindex,hyperfigures,pagebackref=false,a4paper]{hyperref}
\usepackage{latexsym}
\usepackage{mathcomp}
\usepackage{textcomp}
\usepackage{verbatim}
\usepackage{dcolumn}
\usepackage{ulem}

\providecommand{\by}{$\times$}
 
\providecommand{\degr}{$^{\circ}$}

\providecommand{\STO}{SrTiO$_{3}$}

\providecommand{\Vn}[1]{$V^{0}_{\mathrm{#1}}$}

\definecolor{red}{rgb}{0.8,0,0.2}

\def\beeq{\begin{equation}}
\def\eneq{\end{equation}}
\def\beeqa{\begin{eqnarray}}
\def\eneqa{\end{eqnarray}}
\def\Ang{\AA}

\definecolor{adobe}{rgb}{.8,.6,.5}
\definecolor{mygreen}{rgb}{0.0,0.6,0.0}
\definecolor{blue2}{rgb}{0.0,0.0,0.8}
\definecolor{brown}{rgb}{0.6,0.3,0.0}
\definecolor{forest}{rgb}{0.0,0.4,0.0}
\definecolor{grass}{rgb}{0.0,0.55,0.25}
\definecolor{grass2}{rgb}{0.0,0.6,0.25}
\definecolor{gray}{rgb}{0.4,0.4,0.4}
\definecolor{grayish}{rgb}{0.2,0.2,0.4}
\definecolor{khaki}{rgb}{0.9,0.9,0.7}
\definecolor{lightteal}{rgb}{0.0,0.6,0.6}
\definecolor{lightyellow}{rgb}{1.0,1.0,0.5}
\definecolor{maroon}{rgb}{0.7,0.1,0.2}
\definecolor{navy}{rgb}{0.0,0.1,0.7}
\definecolor{olive}{rgb}{0.4,0.4,0.0}
\definecolor{orange}{rgb}{0.9,0.45,0.0}
\definecolor{peach}{rgb}{1.0,.8,.7}
\definecolor{purple}{rgb}{0.4,0,0.55}
\definecolor{teal}{rgb}{0.0,0.5,0.4}
\definecolor{turq}{rgb}{0.3,0.6,0.9}
\definecolor{violet}{rgb}{0.75,0,0.75}
\hypersetup{colorlinks=true, linkcolor=navy, citecolor=blue, urlcolor=blue, filecolor=red, anchorcolor=blue}

\def\FE{\textcolor{black}}

\begin{document}
\DeclareGraphicsExtensions{.ps,.pdf,.eps,png}

\preprint{UPDATED: {\color{maroon} \today}}

\title{{\color{navy} Neutral  Defects in \STO \\Studied with Screened Hybrid Density Functional Theory}}

\author{\firstname{Fedwa} \surname{El-Mellouhi}}
   \email{fadwa.el\_mellouhi@qatar.tamu.edu}
     \affiliation{Chemistry Department, Texas A\&M University at Qatar, Texas A\&M Engineering Building, Education City, Doha, Qatar}

\author{\firstname{Edward N.} \surname{Brothers}}
   \email{ed.brothers@qatar.tamu.edu}
     \affiliation{Chemistry Department, Texas A\&M University  at Qatar, Texas A\&M Engineering Building, Education City, Doha, Qatar}

\author{\firstname{Melissa J.} \surname{Lucero}}
  \affiliation{Department of Chemistry, Rice University, Houston, Texas 77005-1892}

\author{Gustavo E. Scuseria}
  \affiliation{Department of Chemistry, Rice University, Houston, Texas 77005-1892}
  \affiliation{Department of Physics and Astronomy, Rice University, Houston, Texas 77005-1892}
 \affiliation{Chemistry Department, Faculty of Science, King Abdulaziz University, Jeddah 21589, Saudi Arabia }

\begin{abstract}
The properties of neutral defects in \STO~are calculated using the screened
hybrid density functional of Heyd, Scuseria, and Ernzerhof (HSE). The formation
energies, the crystal field splittings affecting the \STO~band structure, as well as the relaxation geometries around each defect are discussed.
Oxygen vacancies  introduced in \STO~ are found to cause a small  tetragonal elongation of the lattice along the $z$-axis. The resulting  conduction band minimum electron effective masses deviate from the bulk values and support the proposal of  enhanced electron mobility along the direction of the compressive strain. The locations of the various defect bands within the  \STO~gap are estimated
without introducing any post-hoc corrections, thus allowing  a more reliable comparison with
experiment. 
\end{abstract}


\maketitle


\section{Introduction}

The properties of point defects in metal oxides and their superlattices have
been an area of recent interest, and have been investigated using increasingly
precise and sophisticated experimental methods.\cite{Liu:2011,Hwang:2012}
Optical experiments can determine differences between sharp photoluminescence
peaks with meV precision;\cite{Ravichandran:2011} this can be used to determine
the different \textit{types} of defects, but the chemical identity of a point
defect or defect complexes remains difficult to obtain.  Electronic structure
methods have been used to obtain the formation energies and the location of
defect levels within the gap to identify the source the photoluminescence
peaks.\cite{Chakhalian:2012,Chakhalian:2011}  However, first principles
calculation of defects in complex systems suffer from several
limitations.\cite{Lambrecht:2011} The long-range interaction between defects in
neighboring cells, an artifact of periodic boundary conditions (PBC), can affect
relaxation around a defect.~\cite{Castleton:2009} Also, correcting the spurious interaction between
charged defects\cite{Makov:1995, Schultz:2006} and the introduced compensating
background charge is an area of ongoing interest.  Furthermore,  there is a
well-known underestimation of band gaps when calculations using the local spin
density approximation\cite{LSDA} (LSDA) of density functional theory (DFT) are
performed; while it is possible to apply a post-hoc correction to the gap and
get good results for uniform solids,\cite{Lany:2008} it is less obvious how
comparable to experiment the result of  this practice is when applied to the
location of the defect levels.  More advanced semilocal density functionals
perform better but still generally underestimate the band gap.~\cite{Deak:2011}

Some of these limitations may be overcome via post-DFT or hybrid DFT
calculations, which often give band gaps closer to experiment.~\cite{Deak:2011}
Band gap accuracy generally implies efficacy in modeling defect structures; for
example, a recent comparison of semilocal and hybrid density functional theory (DFT)
calculations showed that there is a  strong correlation between the calculated
valence band width (VBW) and defect formation energies.\cite{Ramprasad:2012}
Accurate VBW values obtained with the Heyd-Scuseria-Ernzerhof screened hybrid
functional\cite{HSEh} (HSE)  lead to point defect  formation energies and energy
levels in close agreement with experiment for elemental and binary
nonmetals.~\cite{Janoti:2010,Komsa:2010,Komsa:2011}
These findings are encouraging because (unlike semilocal functionals) screened
hybrid functionals require no post-DFT correction of the band gap and provide
excellent lattice parameters in addition to other bulk
properties.\cite{Henderson:2011, Janesko:2009, HISS2, Lucero:2012} In studies of
cubic \STO (STO)\cite{El-Mellouhi:2011, Janotti:2011, Jalan:2011, Wahl:2008} and
other pervoskites,\cite{El-Mellouhi:2012, Hong:2010, Evarestov:2011,
Janotti:2011} HSE has been shown to perform well for band gaps, spin-phonon
effects, and numerous structural effects, including strain, bulk moduli, octahedral angles tilts and rotations
amongst others.  Recently, a regular and hybrid density functional study on many perovskites~\cite{Garcia:2012}  has highlighted  the relevance of anharmonic corrections to lower the octahedral rotation and tilt angles. The  anharmonic  correction was 0.15\degr~for all compounds except STO for which the correction was  0.8\degr. For methods containing 25\% HF mixing, namely PBE0 and PBEsol0,   the uncorrected angles in  \STO~were already  close to experiment; applying the anharmonic correction lead to too small angles. HSE belongs to the same family but has a screening parameter $\omega$ = 0.11, so it can be thought of as an interpolation between PBE ($\omega$ =$\infty$) and the global hybrid PBE0 ($\omega$ =0). We expect the anharmonic effects correction would lead to HSE octahedral rotations overcorrected to too small angles compared to experiment.

We here present the results of our HSE calculations for the formation
energies and band structures of neutral defects in \STO. This work is motivated
by HSE's agreement with experiment in  the calculation of  many of the
electronic, structural and elastic properties in cubic
\STO\cite{El-Mellouhi:2011, Janotti:2011} and other metal
oxides.\cite{El-Mellouhi:2012} It is worth noting that the  direct and indirect
band gaps as well as VBW are in excellent agreement with experiment (see
table~\ref{tab:VBW}) compared to the results from LSDA and
B3PW~\cite{Zhukovskii:2009}, which were previously used to study point defects
in \STO. We focus on the various neutral vacancies in \STO, such as the oxygen
(\Vn{O}), strontium (\Vn{Sr}) and titanium vacancies (\Vn{Ti}), as well as on the
effect of doping STO with La (La$_{Sr}$). 

%
\section{Computational Methods} 

All calculations included in this manuscript were performed using the
development version of the {\sc gaussian} suite of programs,\cite{gdv} with periodic boundary conditions (PBC)\cite{Kudin:2000} used throughout.   The functionals applied in this work include  the
generalized gradient approximation of Perdew, Burke and
Ernzerhof\cite{Perdew:1996,Perdew:1997} (PBE) and HSE.\cite{HSEh}
The Def2-\cite{Weigend:2005} series of Gaussian basis sets were optimized following
our procedure, described in Ref.~\onlinecite{El-Mellouhi:2011}.  We use the
notation SZVP to differentiate these optimized PBC basis sets from the molecular
Def2-SZVP basis sets. Strontium has the inner-shell electrons replaced with smallcore
pseudopotentials, while for oxygen  and titanium  atoms we utilize all
electron basis sets. SZVP basis offers a good compromise between computational efficiency and   high accuracy for electronic and structural properties of bulk \STO.~\cite{El-Mellouhi:2011}   The use of SZVP basis set with HSE screened hybrid functional (HSE/SZVP)
imposes limitations on the size of the supercell that can be efficiently
computed, so a STO supercell of 2$\times$2$\times$2 (40 atoms)  was used with a
dense $k$-point mesh of 6$\times$6$\times$6  which included the $\Gamma$ point.
Calculations with larger supercells of  2$\times$3$\times$3 (90 atoms) with the
same density of $k$-points were performed in order to discuss  the importance of
defect self-interactions and the effect of varying the defect concentration on
the electronic properties of STO.   The pruned integration grid for
DFT employed was (99,590), which corresponds to the
Gaussian option "ultrafine". The remaining numerical settings in {\sc gaussian}
were left at the default values, \textit{e.g.}, geometry optimization threshold was set to  450 $\times$ 10$^{-6}$ hartrees/bohr,
 SCF convergence was set to "tight". Unless
otherwise noted, crystal structures for chemical potential calculations on SrO,
TiO, Ti$_2$O$_3$, and Ti$_2$O were downloaded as CIF files from the ICSD.\cite{ICSD}
All structures are fully relaxed (optimized) unless otherwise noted.  

\begin{table}[!htb]
\caption{ Comparison of properties of bulk
STO relevant to defect formation from our work, previous computational studies,
and experiment. Our HSE/SZVP band gaps and valence band widths (VBW) are
closer to experiment than the semilocal LSDA and the global hybrid B3PW results. }
\begin{ruledtabular}
\begin{tabular}{lllllll}
\label{tab:VBW}
					&LSDA\footnotemark[1] &B3PW\footnotemark[1] &HSE\footnotemark[2]&Exp.\\
\hline
\\
Direct gap (eV)   		&2.36      &3.96   &3.59     	&3.75\footnotemark[3]\\
Indirect gap (eV)   	&2.04      &3.63   &3.20     	&3.25\footnotemark[3] \\
VBW (R$\rightarrow$R)(eV) &4.77 &6.47 	&5.0 		&5.0\footnotemark[3] \\
a$_0$($\text{\Ang}$)   &3.86   &3.90  &3.89   	 	&3.89\footnotemark[4], 3.90\footnotemark[5]\\
B(GPa)	&214 &177 &204 &179\\
\end{tabular}
\end{ruledtabular}
\footnotetext[1] {Ref.~\onlinecite{Piskunov:2004}}\footnotetext[2] {Ref.~\onlinecite{El-Mellouhi:2011}}
\footnotetext[3] {Ref.~\onlinecite{Benthem:2001}}\footnotetext[4]{Ref.~\onlinecite{Abra95}.}
\footnotetext[5]{Ref.~\onlinecite{Hell69}.}
\end{table}
\section{Results and Discussion}
\subsection{Vacancy Formation Energy Calculations }

The calculations of neutral defect formation energies are based upon the formalism
of Zhang and Northrup.\cite{Zhang:1993} These values are generated using the equation
\begin{equation}\label{eqn:tanaka}
    \begin{split}
    E_{f} & = E_{T}-[E_{T}(\mathrm{perfect})\\
        & - ~n_{\scriptscriptstyle{Sr}}\mu_{\scriptscriptstyle{Sr}}-n_{\scriptscriptstyle{Ti}}\mu_{\scriptscriptstyle{Ti}}-n_{\scriptscriptstyle{O}}\mu_{\scriptscriptstyle{O}}]\\
    \end{split}    
\end{equation}

\noindent where $E_{T}$ and $E_{T}(\mathrm{perfect})$ are the calculated total
energies of the supercells containing the point defect and the perfect bulk host
materials, respectively.  The number of each element removed from the perfect
supercell is represented by $n_{\scriptscriptstyle{x}}$,  while
$\mu_{\scriptscriptstyle{x}}$ corresponds to the atomic chemical potentials in an
\STO~ crystal.  Given the assumption that \STO~is always stable, the chemical
potentials of the these elements can vary with the following restriction:

\begin{equation}
\mu_{Sr}+\mu_{Ti}+3\mu_O =\mu_{SrTiO_3(bulk)}
\end{equation}

Atomic chemical potentials vary according to the sample composition and cannot be
determined exactly.  However, they can be varied to cover the whole phase diagram
of \STO, by splitting into  SrO, TiO, Ti$_2$O$_3$, and Ti$_2$O bulk phases.
Hence, the calculated  formation energies for the  neutral point defects vary
according to equilibrium positions; an example of this is O-rich versus O-poor
points on the phase diagram.  The calculated enthalpies of formation in idealized materials (non-relaxed structures)
for phases containing Sr, Ti and O are summarized in Table~\ref{tab:enthalpies},
and are compared to previous LSDA calculations\cite{Tanaka:2003} and
experiment.\cite{NIST:2003} 

\begin{table}[!ht]
\caption{\label{tab:enthalpies} Calculated enthalpies of formation in eV/atom for idealized materials for phases containing Sr, Ti and O  compared to previous LSDA calculations and experiment.}
\begin{ruledtabular}
\begin{tabular}{lllllll}
        &HSE &PBE &LSDA\footnotemark[1] &Exp.\footnotemark[2]\\
\hline
\vspace {-2mm}
\\
TiO$_2$    &-3.92 &-3.75    &-3.76    &-3.24\\
Ti$_2$O$_3$&-3.74 &-3.60    &-3.63    &-3.15\\
TiO        &-2.99 &-2.95    &-3.04    &-2.81\\
SrO         &-4.00&-3.90 &-3.36     &-3.07\\
\end{tabular}
\end{ruledtabular}
\footnotetext[1]{Ref. \onlinecite{Tanaka:2003}} 
\footnotetext[2]{Ref. \onlinecite{NIST:2003}}
\end{table}
%
As a general trend, the formation enthalpies computed with HSE are close to  the
results from semilocal functionals LSDA and  PBE (from this
work), although the HSE values are
slightly higher. The only exception is SrO, where both HSE and PBE tend to
overestimate the formation enthalpies to the same extent, exceeding the LSDA
values. 
%
\begin{figure}[!h]
\includegraphics[width=0.8\columnwidth]{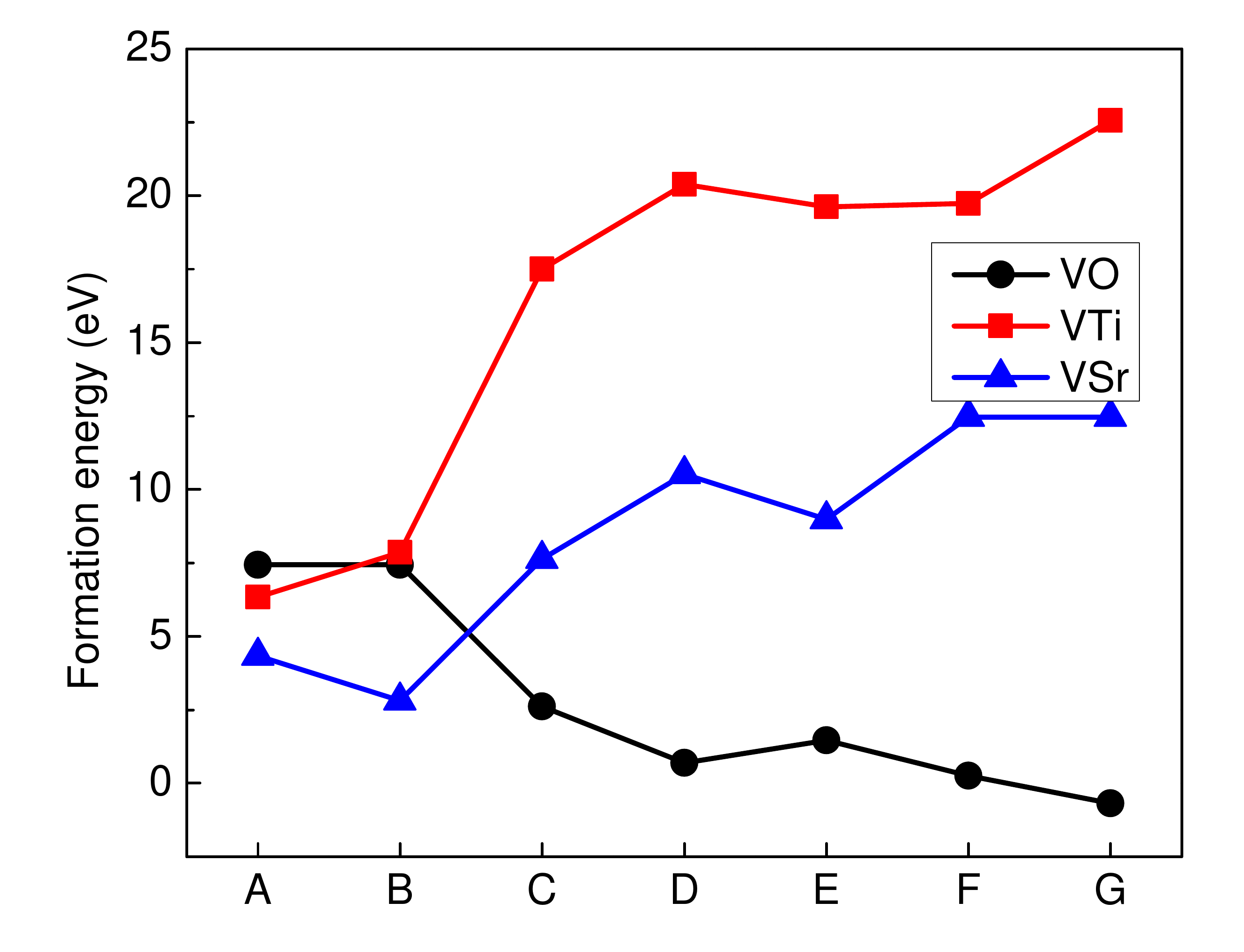}
\caption{\label{fig:formation} (Color online) Defect formation energies of isolated neutral vacancies in STO at each equilibrium point based upon the phase diagram detailed in the text.}
\end{figure}

The formation energies of vacancy defects in STO as function of its composition
are plotted in Figure~\ref{fig:formation} with the points A to G 
\footnote{ Point A: $\mathrm{\mu_O=\mu_O(bulk)}$,$\mathrm{ \mu_{Sr}+\mu_{O}=\mu_{SrO}(bulk)}$ where $\mathrm{\mu_O(bulk)}$ corresponding to the chemical potential per atom of O$_2$ gas.\\
Point B: $\mathrm{\mu_O=\mu_O(bulk)}$ , $\mathrm{ \mu_{Ti}+2\mu_{O}=\mu_{TiO_2}(bulk)}$. \\
Point C: $\mathrm{\mu_{Ti}+2\mu_{O}=\mu_{TiO_2}(bulk)}$ , $\mathrm{ 2\mu_{Ti}+3\mu_{O}=\mu_{Ti_2O_3}(bulk).}$ \\
Point D:   $\mathrm{\mu_{Ti}+\mu_{O}=\mu_{TiO}(bulk)}$ , $\mathrm{ 2\mu_{Ti}+3\mu_{O}=\mu_{Ti_2O_3}(bulk)}$\\
Point E:   $\mathrm{\mu_{Ti}+\mu_{O}=\mu_{TiO}(bulk)}$, $\mathrm{\mu_{Ti}=\mu_{Ti}(bulk)}$\\
Point F:  $\mathrm{\mu_{Ti}=\mu_{Ti}(bulk)}$ , $\mathrm{\mu_{Sr}=\mu_{Sr}(bulk)}$\\
Point G:  $\mathrm{\mu_{Sr}=\mu_{Sr}(bulk)}$ , $\mathrm{\mu_{Sr}  + \mu_{O}=\mu_{SrO}(bulk)}$} 
based on the phase diagram in  Ref.~\onlinecite{Tanaka:2003}. The plot shows that
HSE's formation energies show a  similar dependence on the chemical potentials
as was observed for the LSDA results of Ref. \onlinecite{Tanaka:2003}.  However, both
\Vn{Ti} and \Vn{Sr} curves  are shifted to higher formation energies, while the
\Vn{O} curve is slightly shifted downwards compared to the LSDA spectrum.  At
point A,  \Vn{Sr} has the lowest formation energy followed by \Vn{Ti} and \Vn{O}.
At point  B, \Vn{Sr} remains the most stable but  \Vn{O} becomes more stable than
\Vn{Ti} contrary to the LSDA predictions. Under  low partial pressures of oxygen
(O-poor conditions corresponding to points $C$ to $G$),  \Vn{O} vacancies form
more easily and dominate the spectrum while the formation energies of  \Vn{Ti} and
\Vn{Sr} keep increasing.

A quantitative comparison  with previously published calculations using the same
cell size formation energies~\cite{Carrasco:2006,Zhukovskii:2009,Tanaka:2003}
available at  point  B (oxygen rich conditions)  reveal several issues of interest.

\begin{itemize}
\item For \Vn{O}, the semilocal functional  PW91\cite{Carrasco:2006} yields a
formation energy of 8.56~eV,  the global hybrid B3PW~\cite{Zhukovskii:2009}
predicted a higher formation energy  of 8.74~eV, while LSDA~\cite{Tanaka:2003}
gave a lower value at 7.95~eV. Our HSE value is the lowest among these, yielding a
formation energy of  7.43~eV. This decrease in the \Vn{O} formation energy is one of the factors leading to a strong competition with  \Vn{Ti} (see below).
\item For \Vn{Ti}, the semilocal functional  LSDA~\cite{Tanaka:2003} yields a
formation energy of 5.7~eV making it far more stable than \Vn{O}. In contrast, with  HSE we predict that  $E_{f}=7.86$ eV  meaning that \Vn{Ti} becomes less stable than \Vn{O}.
\item   \Vn{Sr}  remains the most stable defect under these conditions, with HSE
providing a value of 2.81~eV compared to the  1.7~eV from LSDA.
\end{itemize}

Overall, the quantitative formation energies differences between our HSE results and previous LSDA results are substantial. These might originate from the enhanced accuracy we gain with HSE in the calculated valence band width (VBW) and band gap of bulk \STO~(see table~\ref{tab:VBW}) identified in Ref. \onlinecite{Ramprasad:2012} to reflect an enhanced precision in the defect formation energies.

It is also worth mentioning size effects.  The formation energy of \Vn{O} at
point B in a larger supercell of 90 atoms, giving a 1.8\%  \Vn{O}  concentration,
is  7.66~eV,  which is 0.2~eV higher than the value obtained with the smaller
supercell. 


%
    \begin{figure}[!htb]
     \includegraphics[width=0.5\columnwidth]{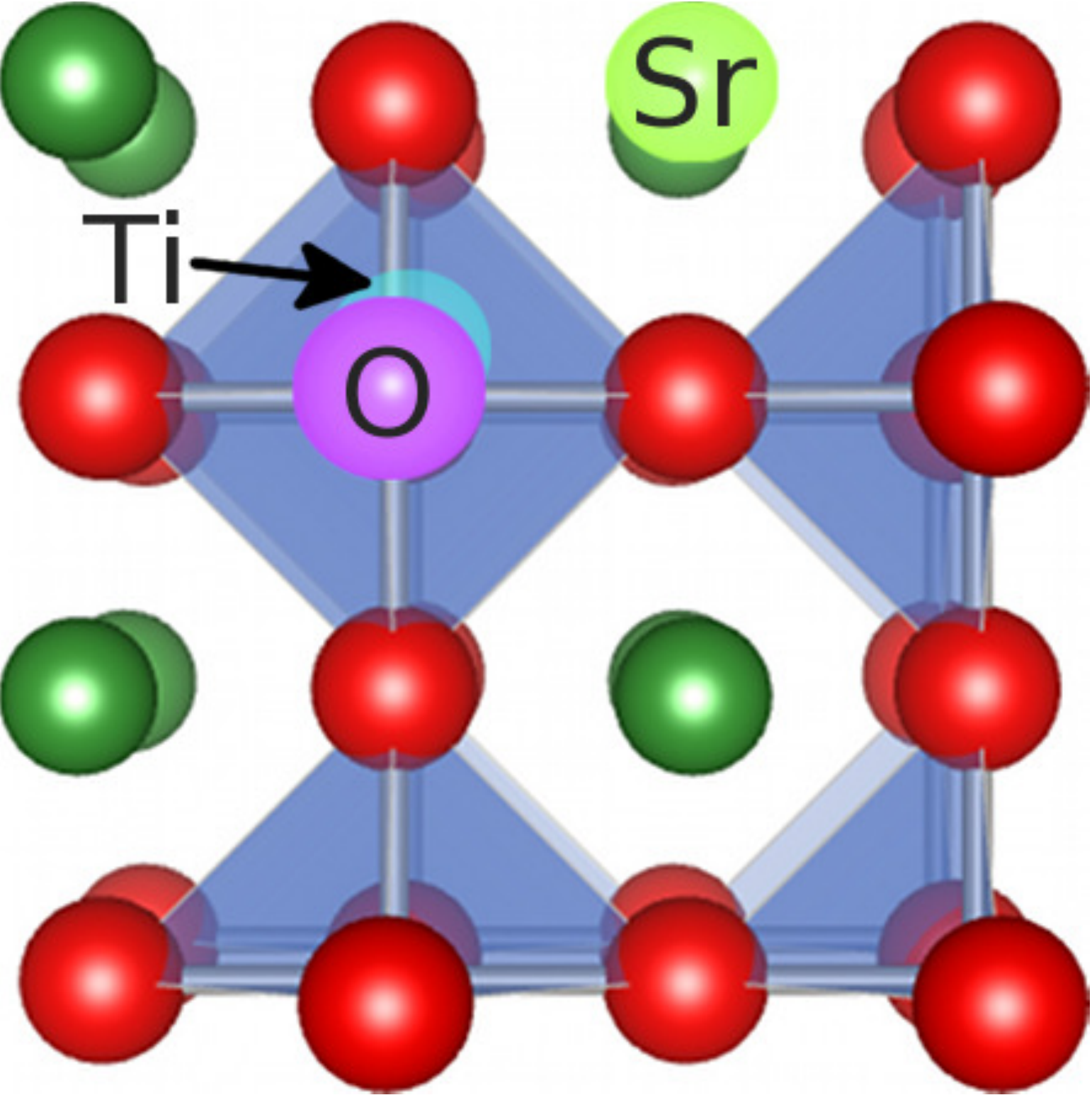} 
    \caption{\label{fig:allvacancies}(Color online). Ideal, crystalline \STO, as
observed from the 100 face, where the \textcolor{red}{oxygen} anions are in
\textcolor{red}{red}, the \textcolor{grass}{strontium} cations are in
\textcolor{grass}{green} and the \textcolor{blue}{titanium} cations are
in \textcolor{blue}{blue}. \textcolor{magenta}{Magenta},
\textcolor{green}{yellowish} and \textcolor{cyan}{cyan} circles highlight the
locations of the isolated \textcolor{magenta}{\Vn{O}}, \textcolor{green}{\Vn{Sr}}
and \textcolor{cyan}{\Vn{Ti}} defects, respectively.} 
   \end{figure}
 \subsection{Isolated Vacancies: Relaxation Effects}
\label{sec:relax}
\begin{figure*}
    \includegraphics[width=0.5\columnwidth]{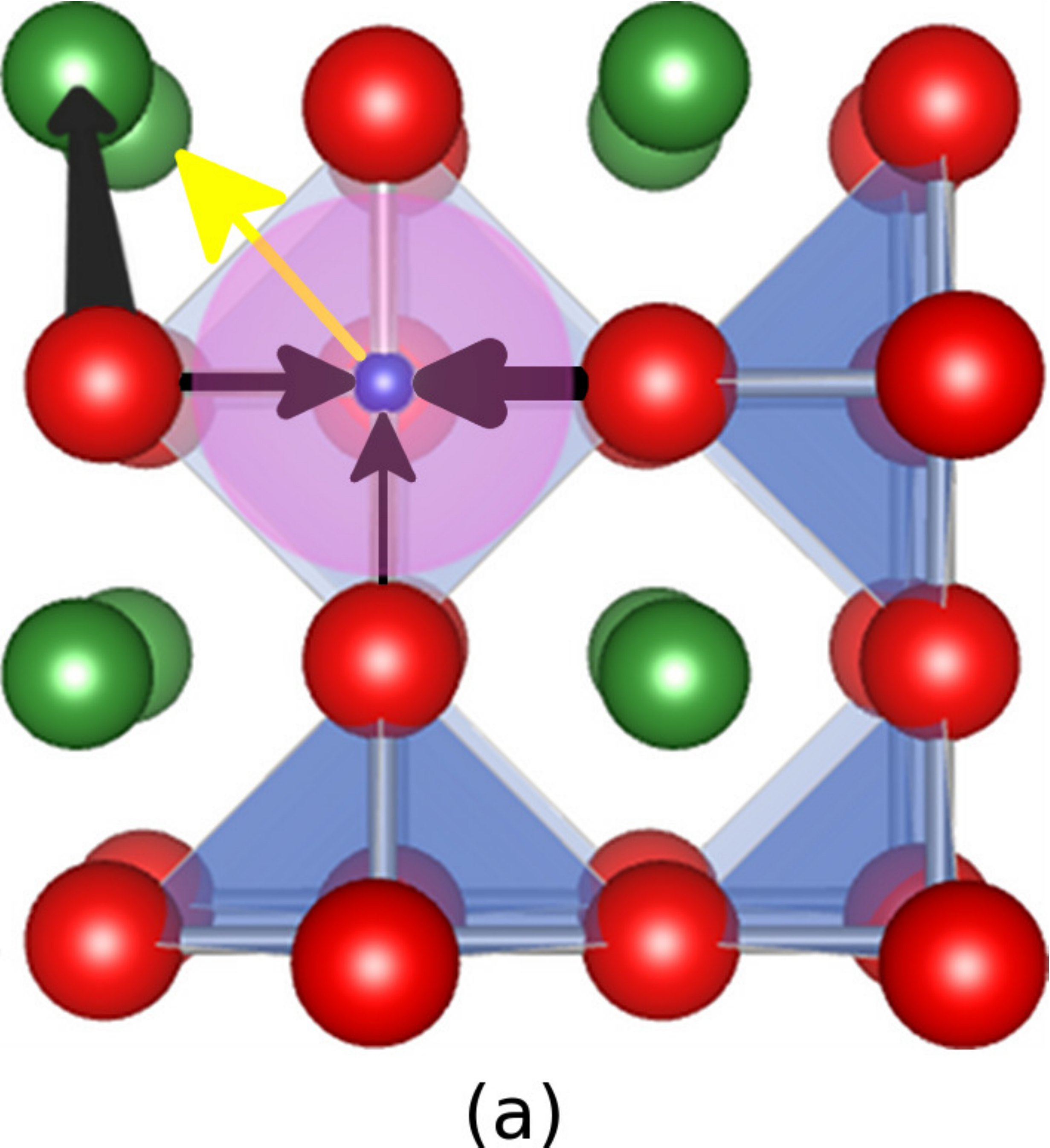}\vspace{5mm}
    \includegraphics[width=0.5\columnwidth]{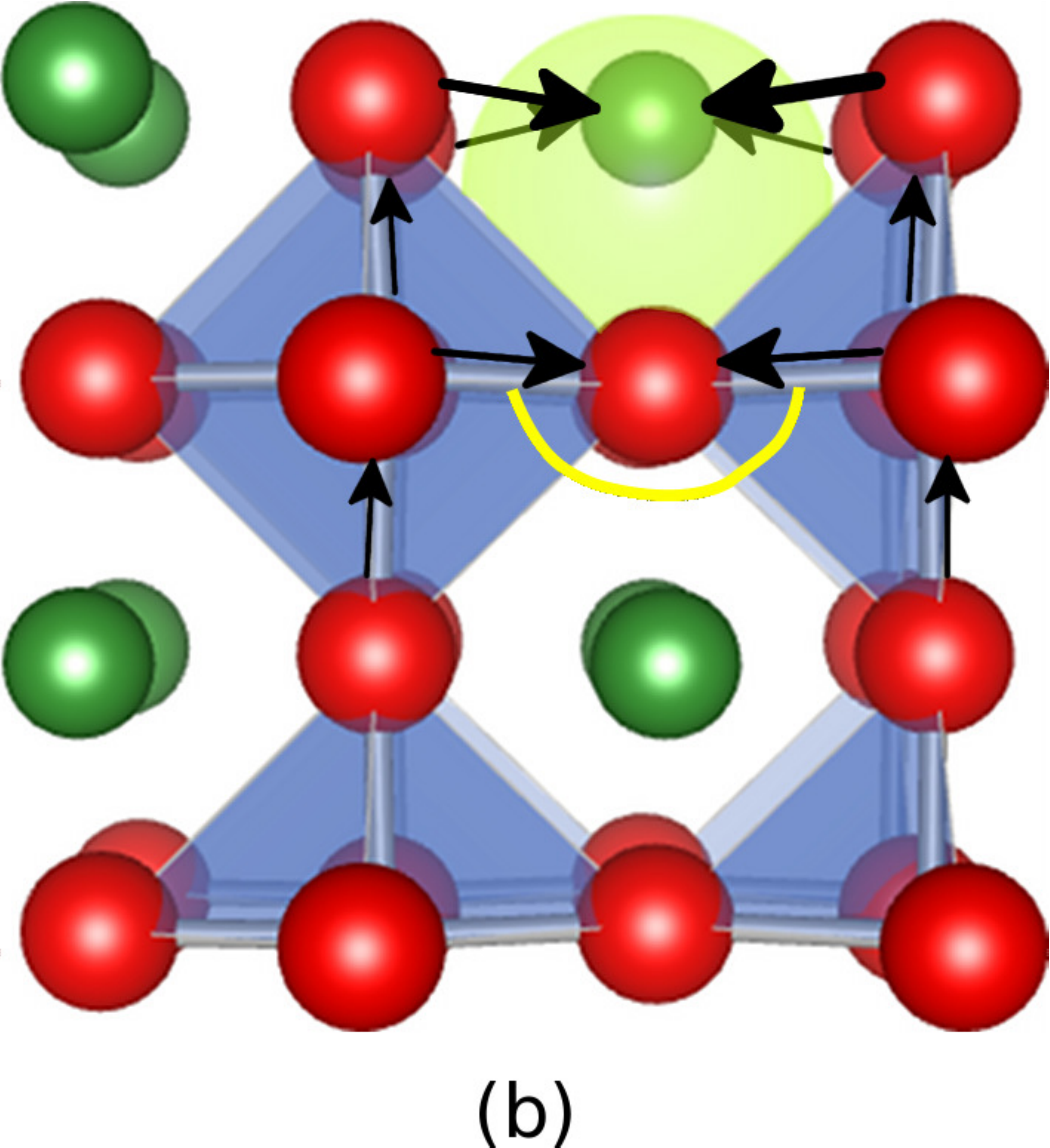}\vspace {5mm}
     \includegraphics[width=0.5\columnwidth]{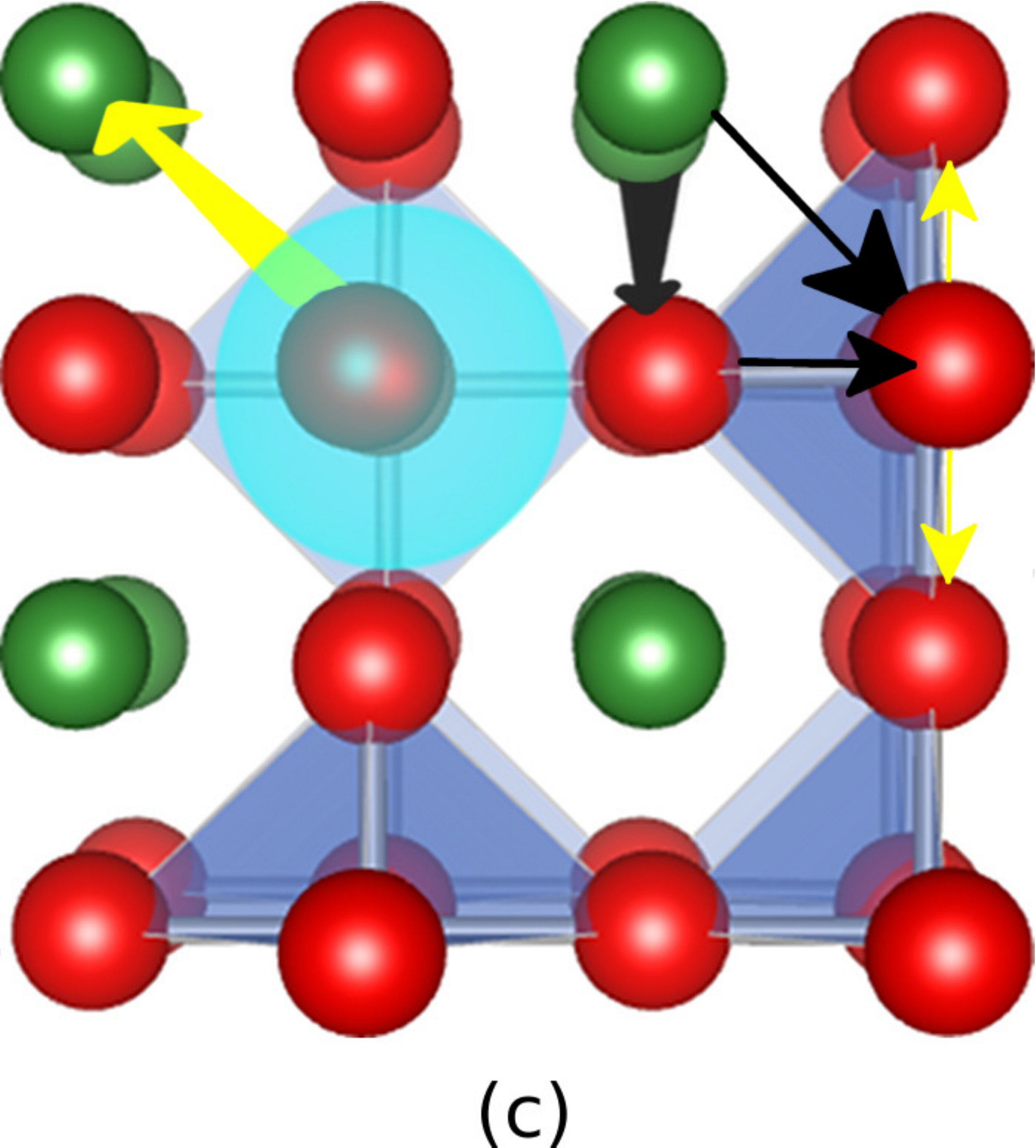} 
    \caption{\label{fig:vacancies}(Color online). Relaxed \STO ~with a single
isolated vacancy. Each structure is in the 100 orientation and the location of the
vacancy is noted.   The distortions induced by \Vn{O}, \Vn{Sr} and \Vn{Ti}  are
highlighted(a), (b) and (c), respectively.  Oxygen anions are red, Sr cations are
green, and Ti cations are blue.}
    \end{figure*}

 %
Isolated, neutral defects have been introduced into the crystal structure of cubic
STO by removing one atom of either O, Sr or Ti, respectively, as depicted in
Figure~\ref{fig:allvacancies}.  The structure was then fully relaxed using HSE/SZVP.
Figure~\ref{fig:vacancies} illustrates the major displacements in each of the
defect structures, namely \Vn{O}, \Vn{Sr} and \Vn{Ti}, relative to the idealized crystal.
Distance \textit{decreases} are depicted by black arrows, while
\textit{increasing} lengths are denoted by yellow arrows.  In each case, the
magnitude of displacement is implied by the size of the arrows: larger/thicker
arrows  indicate greater deviation from the defect-free structure.
%
%

%
A 4\%  \Vn{O}  vacancy was created by removing one oxygen (Figure ~\ref{fig:vacancies}-a) from a 2\by2\by2 cell, inducing small, asymmetric bond length changes throughout the cell and an overall decrease in volume relative to the crystal.  At the vacancy, the remaining O-Ti bonds all contract but to different extents. Near the vacancy, oxygens move toward nearby strontium, inducing  O-Sr bond contractions of up to 5.66 pm.  Further from the vacancy, in the regions of the cell behind the defect (and the plane of the figure), strontium appears to migrate inward, toward the vacancy, causing the more distant O-Sr bonds to lengthen by as much as 6.05 pm.

The \Vn{O} structure contracts, and undergoes a tetragonal distortion where the lattice parameters are  $a$= 3.879
and  $b$=$c$=3.890~\AA, compared to the 3.902 \AA~in the crystal. High concentration of oxygen vacancies were identified to be responsible of a  similar tetragonal distortion reported experimentally~\cite{Ravichandran:2011} in STO  doped with La. Worth noting that with  HSE/SZVP, we are able to capture this small tetragonal distortion corresponding to a $c/a$ ratio of 1.0028. This distortion  was also observed in previous LDA calculations~\cite{Luo:2004}  for
supercells of the same size, but the $c/a$ ratio was not reported.  

As seen from Figure~\ref{fig:vacancies}-b, the atomic relaxation around the
\Vn{Sr} induce staggered bond lengths (vertically) and a general contraction of O-Sr bonds, with the large decrease being 1.55 pm, smaller than those observed for O-Sr in the presence of an oxygen vacancy.
The alternating distances cause the linear~Ti-O-Ti angle to decrease to 173.9\degree.
Again, the defect structure contracts, with all lattice parameters decreasing:
$a$=$b$=$c$=3.880\AA~.

The \Vn{Ti} structure depicted in Figure~\ref{fig:vacancies}-c shows the axial oxygens experiencing a 9.2 pm decrease in length, while the medial oxygens move away from the vacancy by 0.6 pm. 
The Sr-O distances shrink over a range of 1.51 to 5.40 pm. While also asymmetric, these distortions have less of an effect on the cell., with the equilibrium lattice parameters being $a$=$c$=3.892 and  $b$=3.894
The $a/b$ ratio here is very small so the structure remains effectively cubic following full relaxation.

In summary, the loss of metallic species results in smaller volumes with the loss
of the larger metal, \Vn{Sr}, producing greater contraction.  Removing the
oxygen produces larger deviations from cubic symmetry in addition to overall
shrinkage of the unit cell.

\subsection{Isolated Vacancies: Band Structures}
\begin{figure*}[!htb]
\includegraphics[height=0.24\textheight]{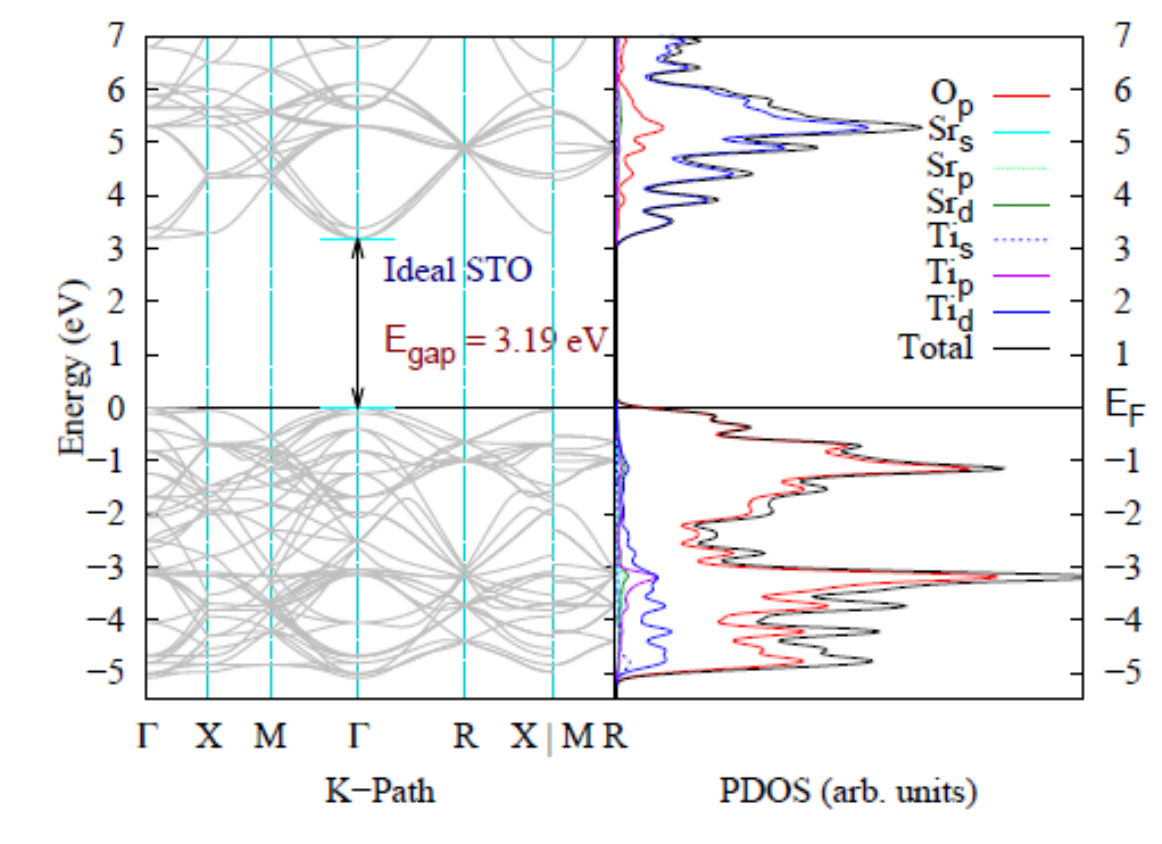}\includegraphics[height=0.24\textheight]{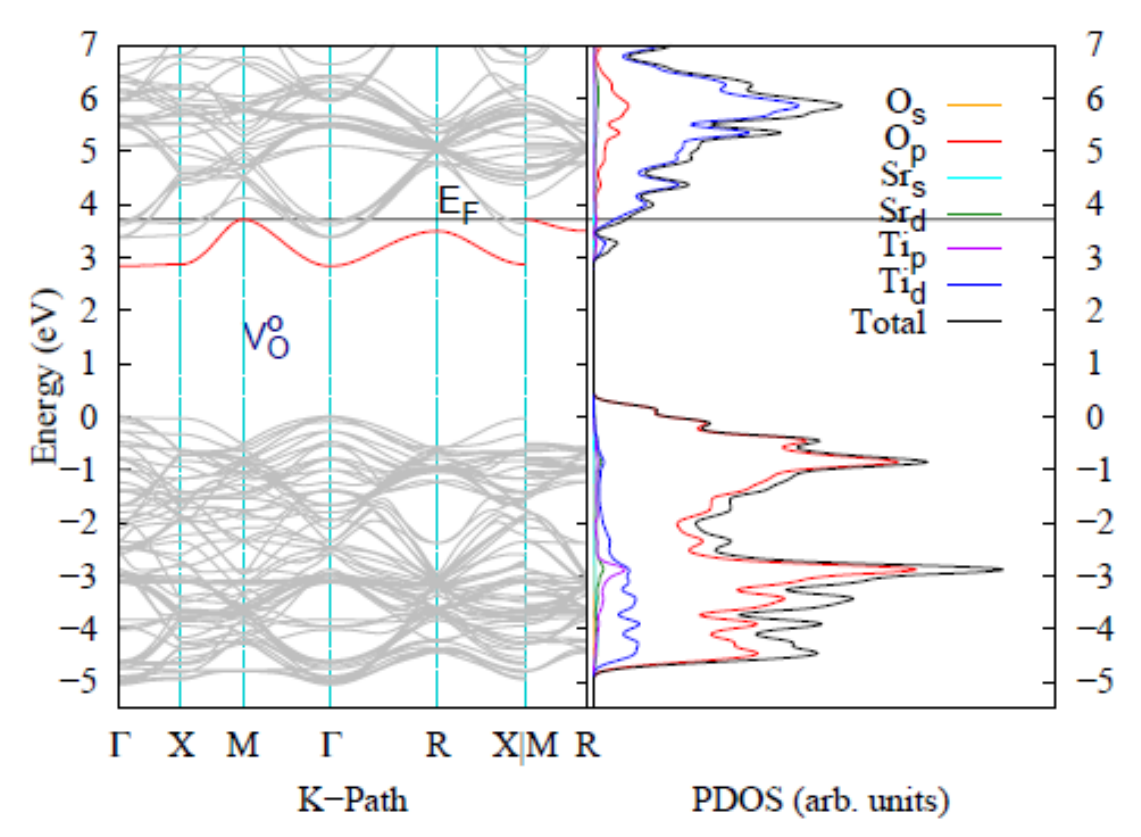}
\includegraphics[height=0.24\textheight]{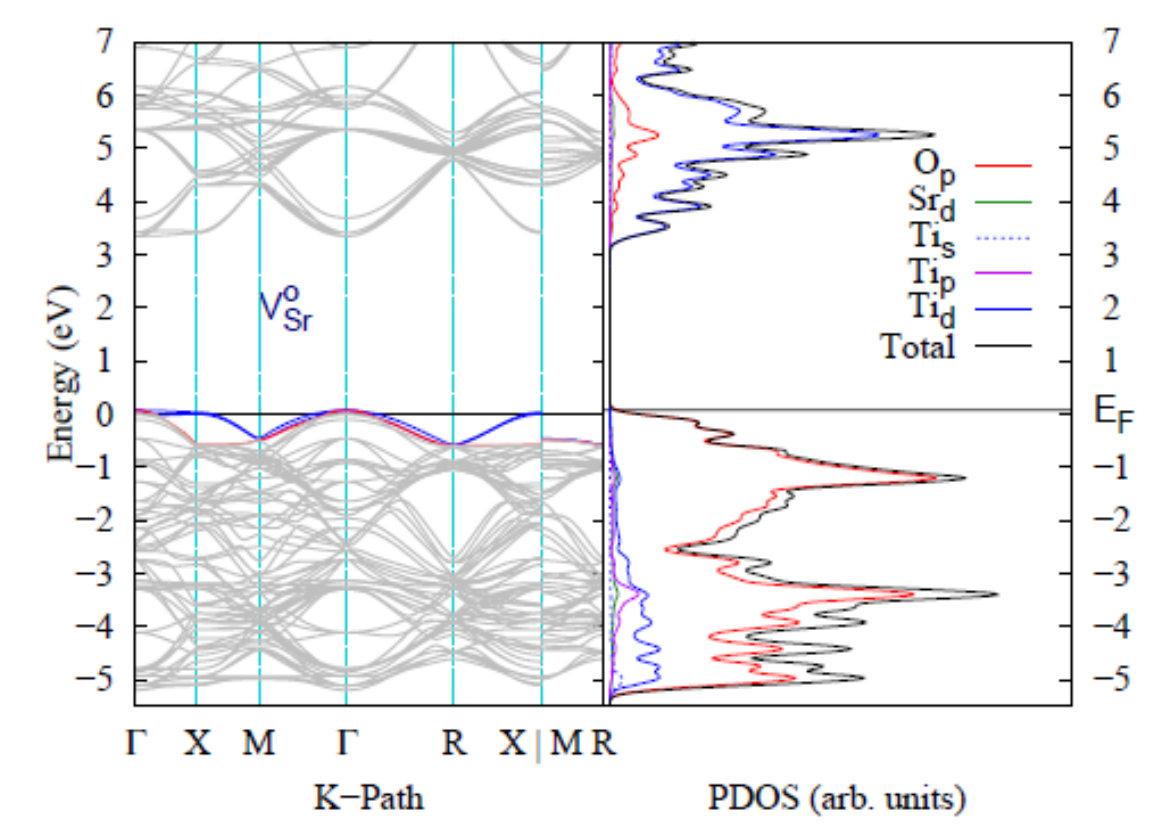} \includegraphics[height=0.24\textheight]{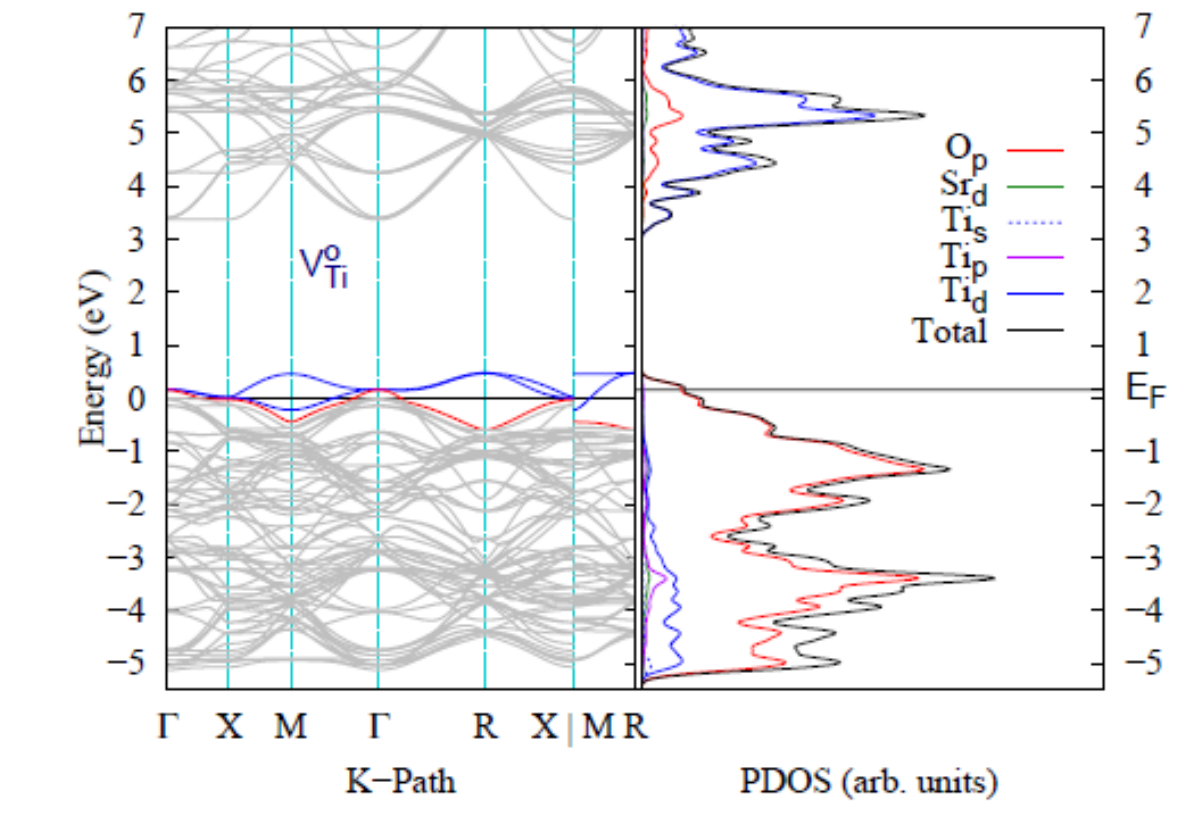}
\caption{\label{fig:combo}(Color online). Band structures and PDOS calculated with HSE/SZVP for the 2\by2\by2 \STO~supercell. The top figures represent bulk STO,  and \Vn{O} while the bottom row contains \Vn{Sr} and \Vn{Ti}. The Fermi  energy \textit{E$_{F}$} is indicated by a solid black line. The red bands indicate the occupied defect bands while the unoccupied defect bands are shown in blue.}
\end{figure*}

The electronic band structure and the projected densities of states (PDOS) of the
ideal and nonideal/defect STO supercells (doped with O, Sr and Ti neutral vacancies)
are shown in figure~\ref{fig:combo}.  To allow a clear comparison between the
different systems, the VBM was set as reference for all systems while the  Fermi
 energy is depicted with a solid line. The PDOS are plotted alongside with their
corresponding band structure and rescaled in the same way.  

For the ideal 2\by2\by2  STO supercell the Brillouin zone folds, causing a direct
band gap of 3.19 eV. The CBM in the ideal solid is triply degenerate at $\Gamma$
and composed of a heavy electron band ($he$),  a light electron band ($le$) and a
spin-orbit band ($so$). The band degeneracy is lifted as we move far from the
$\Gamma$ point towards high symmetry directions $X$, $M$ and $R$, giving rise to
different values of the electron effective masses $m_{he}$, $m_{le}$, $m_{so}$ in
each direction. The electronic effective mass for each band $i$  can be computed
by the curves of energy versus $\vec{k}$ 
being fitted to a parabola~\cite{Wunderlich:2009}; $\vec{k}$ is 
taken from the $\Gamma$ point up to 0.5\% along $\Gamma \rightarrow  X$,
$\Gamma \rightarrow M$ and $\Gamma \rightarrow R$ paths using the formula: 
\begin{equation}\label{eqn:mass}
\frac{1}{m}=\frac{2}{\hbar}\frac{d^2E}{dk^2}
\end{equation}

We estimate the $\Gamma\rightarrow X$ effective masses, without spin-orbit correction, $m^{\Gamma\rightarrow
X}_{he}$ and  $m^{\Gamma\rightarrow X}_{le}$ to be 7.3$m_e$ and 0.5$m_e$
respectively,  where $m_e$ is the free electron mass. This agrees well with
previous nonrelativistic calculations~\cite{Janotti:2011,Wunderlich:2009,Marques:2003}  using LSDA
and HSE, which predicted $he$ masses ranging from 6.1 to 7.3$m_e$ and $le$ masses of
0.4$m_e$. The remaining  effective masses are $m^{\Gamma\rightarrow
M}_{he}$=1.1$m_e$,  $m^{\Gamma\rightarrow M}_{le}$=0.8$m_e$,
$m^{\Gamma\rightarrow R}_{he}$=0.9$m_e$ and  $m^{\Gamma\rightarrow
R}_{le}$=0.7$m_e$. 

%
\begin{table}[!b]
\caption{Location of the various defect bands averaged in the Brillouin zone with respect to the conduction band minimum ( $\Delta\epsilon_{\mathrm{DB-CBM}}$ ) and valence band maximum ($\Delta\epsilon_{\mathrm{DB-VBM}}$).  Data are for the 2$\times$2$\times$2 \STO~supercell computed with HSE/SZVP. Defect formation energies (in eV) at point B (oxidation condition) are reported as well. }
\begin{ruledtabular}
\begin{tabular}{llllll}
\label{tab:band}

		&$\Delta\epsilon_{\mathrm{DB-CBM}}$ &$\Delta\epsilon_{\mathrm{DB-VBM}}$  &E$_f$\\
\hline
\\
\Vn{Ti} 	&3.06	& 0.15	 &7.86\\
\Vn{Sr}  &3.27	& 0.07	  &2.81\\

\Vn{O} 	& 0.44, 0.40\footnotemark[1] 	 &2.82, 2.90\footnotemark[1] 	&7.43\\

\end{tabular}
\end{ruledtabular}
\footnotetext[1]{Experimental  estimations from Refs. ~\onlinecite{Kan:2005,Ravichandran:2011}} 
\end{table}

 Introducing \Vn{Ti} into STO, the defect band averaged in the Brillouin zone is located at 0.13~eV above the valence
band maximum (VBM) and 3.06~eV below the conduction band maximum (CBM) (see
Table~\ref{tab:band}) . The defect band has a VB
character and is triply degenerate  at $\Gamma$ with a single band  occupied by
two electrons, while the upper two bands are  empty (see  figure~\ref{fig:combo}).  The crystal field resulting from the
atomic relaxation around \Vn{Ti}   causes the triply degenerate CBM band to split in
the following way: the heavy electron band ($he$) is followed by the light
electron band ($le$) at 26 meV, while the spin-orbit band ($so$) is located 14 meV
higher.

For \Vn{Sr}, a  triply degenerate defect  band appears  at 0.07~eV above the VBM
and 3.27~eV  below the  CBM  at the $\Gamma$ point. From  figure~\ref{fig:combo}, two bands are occupied by four
electrons while the upper band is empty. The degeneracy of the defect band is
lifted at the X point, where the two occupied band remain degenerate while the
unoccupied band is 600~meV higher in energy. The $he$ and $le$ bands in the CBM
remain degenerate while the $so$ band is located at 1.5 meV higher in energy
mainly because the overall deformation \Vn{Sr} introduces into the lattice is
small.

\begin{figure*}[!htb]
\includegraphics[width=0.40\textwidth]{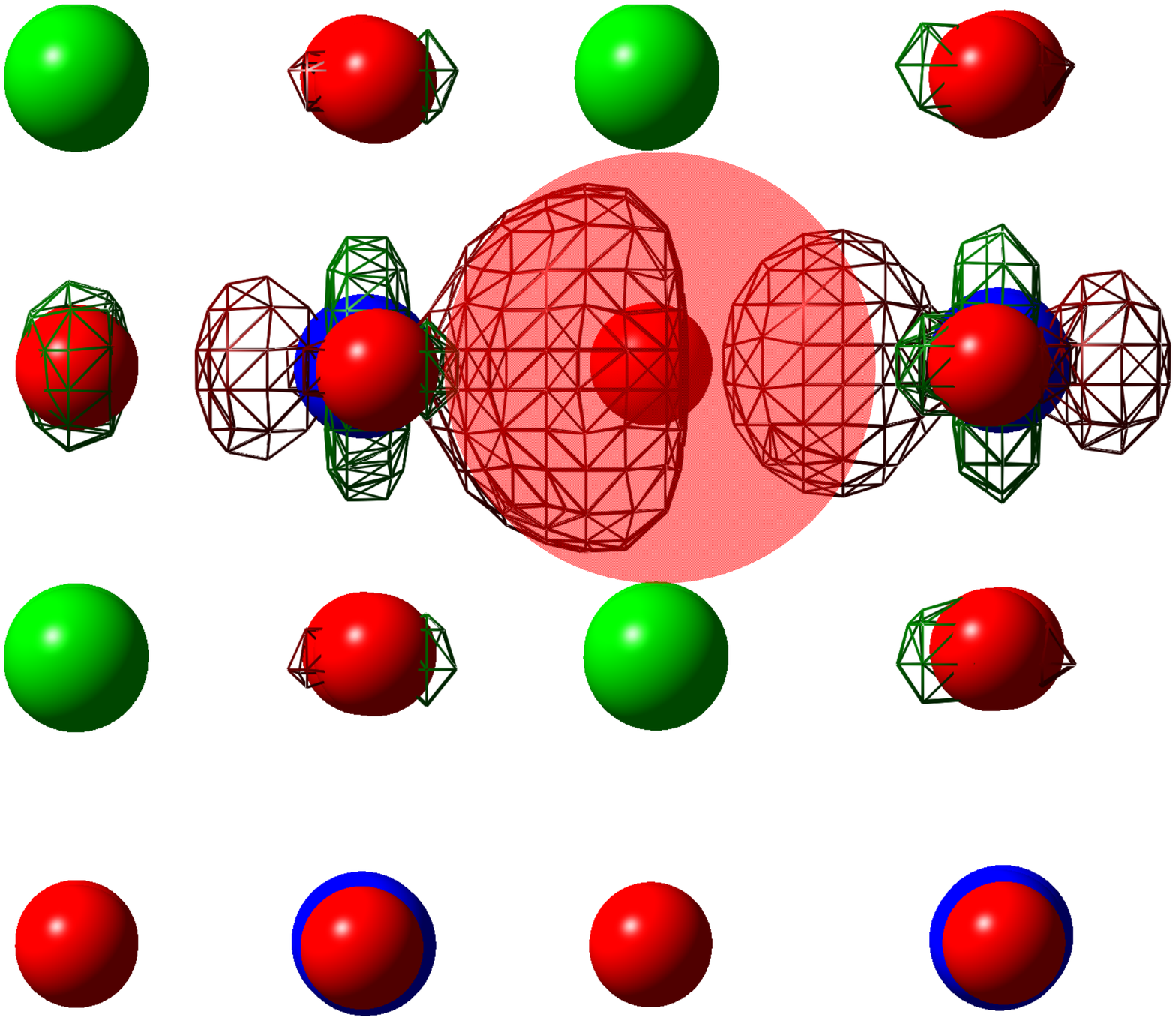}\includegraphics[width=0.45\textwidth]{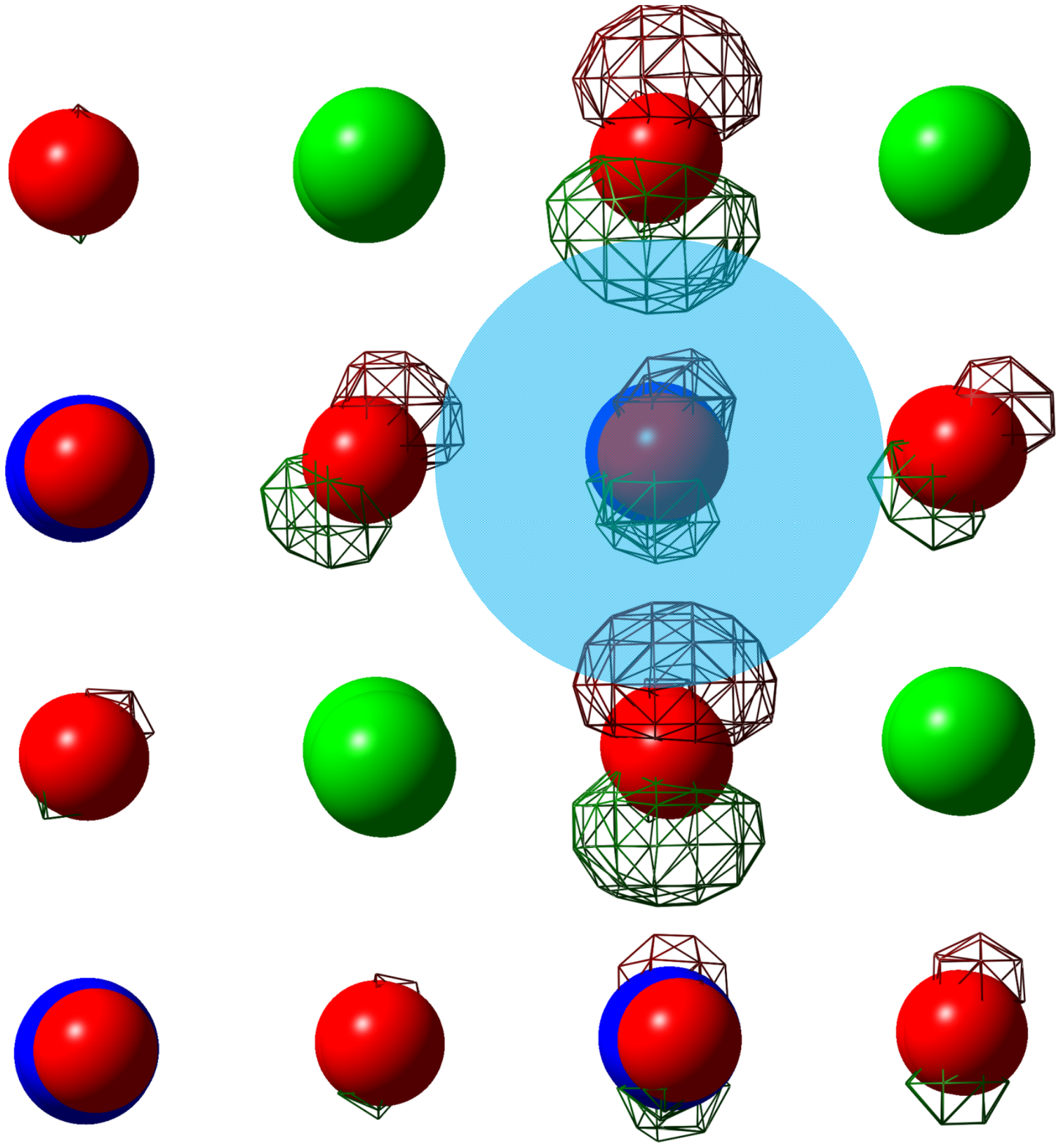}
\caption{\label{fig:iso}Isosurface of the highest occupied  orbitals viewed along the 100 direction showing (a) orbital localization around (a) \Vn{O} (left) \Vn{Ti} (right).  }
\end{figure*}

Introducing \Vn{O} into STO causes  a defect band (DB) populated with two
electrons to appear in the gap just below the bulk-like conduction band maximum labelled here as
(DCBM), indicating  a donor band. The Fermi energy is shifted from the top of the
valence band to the maximum  of the defect band  at the $M$ point followed by the
empty conduction band.  
\FE{The CBM conserves the bulk character as it remains empty and  triply degenerate (clearly shown in Figure ~\ref{fig:combo}) corresponding to the Ti t$_2g$ states followed by a doubly degenerate e$_g$ band. The above CBM conservation of degeneracy confirms that the band appearing underneath is a fully occupied non-degenerate defect band. The isosurface of the highest occupied  state at the $\Gamma$ special k-point (Figure~\ref{fig:iso}) shows that the electronic charge density is  localized on \Vn{O} occupying the  Ti dangling bonds. }

We calculated the position of the defect state by averaging over the Brillouin zone. We found that the DB is located in average at 0.44 eV below the CBM for the 40 atoms supercell  (see table~\ref{tab:band}) and 0.42 eV for the 90 atoms supercell. This could be compared the recent experimental measurements, which place the position of the defect level 0.4~eV below the CBM, causing the blue light photoluminescence of STO at room
temperature.~\cite{Kan:2005,Ravichandran:2011}  In the present calculations,  self-defect interaction or the so-called size effects were minimized by keeping  the defects  neutral, and conserving the same high density of $k$-point sampling of the  Brillouin zone.~\cite{Probert:2003} Nevertheless, simulations using larger supercells and finite size scaling~\cite{Castleton:2009} are still  needed for a more accurate comparison. As of today's computational resources,  a full relaxation of larger supercells using HSE/SZVP and the  computational settings used here are computationally very expensive. This defect band seems to be shallower than the one computed with B3PW, which was located 0.79 eV below the CBM.\cite{Zhukovskii:2009} We
attribute this to the B3PW functional overestimating the band gaps (indirect gap
of 3.6~eV compared to 3.2~eV from experiment). Our HSE-predicted defect level for
\Vn{O} is deeper than that calculated with LDA; in those calculations the defect
band can often not  be distinguished from the CBM (resonant band),~\cite{Luo:2004}
although in some cases it lies as little as  0.08~eV below the
CBM.~\cite{Tanaka:2003} 

The atomic relaxation around \Vn{O} leads to a  tetragonal distortion of the
supercell where  $c/a$=1.0028 (see Section~\ref{sec:relax}), corresponding to a
$-$0.3\% compressive strain along the $X$ axis.  This amount of strain is
comparable to the values applied  experimentally in La doped
STO~\cite{Jalan:2011} and identified as causing a substantial increase of the
electron mobilities (by a factor of 3.3) due to the appearance of light electron
effective mass along the strain and the transport direction.

In our calculation, the effective mass  $m^{\Gamma\rightarrow X}_{he}/m_e$ along the
direction parallel to the  compressive strain ($k^{\parallel}=\pi/a(100)$) drops
from the bulk value of 7.3 to 0.5 in defective \STO (se table~\ref{tab:masses}),
indicating that the electron mobility increases substantially along the x-axis.
However, $m^{\Gamma\rightarrow X}_{he}/m_e$  increases  along the
$k^{\perp}=\pi/a(010)=\pi/a(001)$ directions (perpendicular to the  compressive
strain) to 11.0 indicating a reduced  electron mobility in these directions. Along
$\Gamma\rightarrow M$, the effective massesremain unchanged in the
$k^{\parallel}=\pi/a(110)=\pi/a(101)$  containing the strain axis.

\begin{table}[!b]
\caption{\label{tab:masses}  Non-relativistic HSE/SZVP calculated heavy electron effective masses (in units of the free-electron mass $m_e$) at the  conduction maximum along different high-symmetry directions in ideal ($m_{he}^{CBM}$) and	oxygen vacancy doped  \STO~($m_{he}^{DCBM}$). $k^{\parallel}$ and $k^{\perp}$ represent the  parallel and perpendicular plans to the direction of the compressive $100$ strain resulting from \Vn{O}. }
\begin{ruledtabular}
\begin{tabular}{lll}
										
Direction	& $m_{he}^{CBM}$		&$m_{he}^{DCBM}$  \\		
	\hline
\multicolumn{3}{c}{$\Gamma\rightarrow X$(100) }\\ \\
$k^{\parallel}$ &7.3 &{ 0.5}	\\
$k^{\perp}$ &7.3	&{\ 11.2} \\
\multicolumn{3}{c}{$\Gamma\rightarrow M$(110)} \\ \\

$k^{\parallel}$ &1.1 	&1.1	\\
$k^{\perp}$ &1.1	&1.1	\\
\multicolumn{3}{c}{$\Gamma\rightarrow R$(111)} \\ 
 &0.9	&0.9	\\
\end{tabular}
\end{ruledtabular}

\end{table}


\subsection{STO doped with La}

The substitution of one Sr atom by a La atom in our 2\by2\by2 STO supercell lead
to a dopant concentration of 12.5\%, which is low enough to be compared
to experimental data,\cite{Ravichandran:2011} where dopant concentrations as high
as 15\% has been used. The resulting Sr$_{0.875}$La$_{0.125}$TiO$_3$ compound
relaxes to  a cubic structure with lattice parameters $a$=$b$=$c$=3.893~\AA.
This agrees well with experiments~\cite{Ravichandran:2011} which demonstrated
that La doping conserves the cubic symmetry, and has a negligible effect on the
$c$-axis lattice parameter.

\begin{figure}
\includegraphics[height=0.25\textheight]{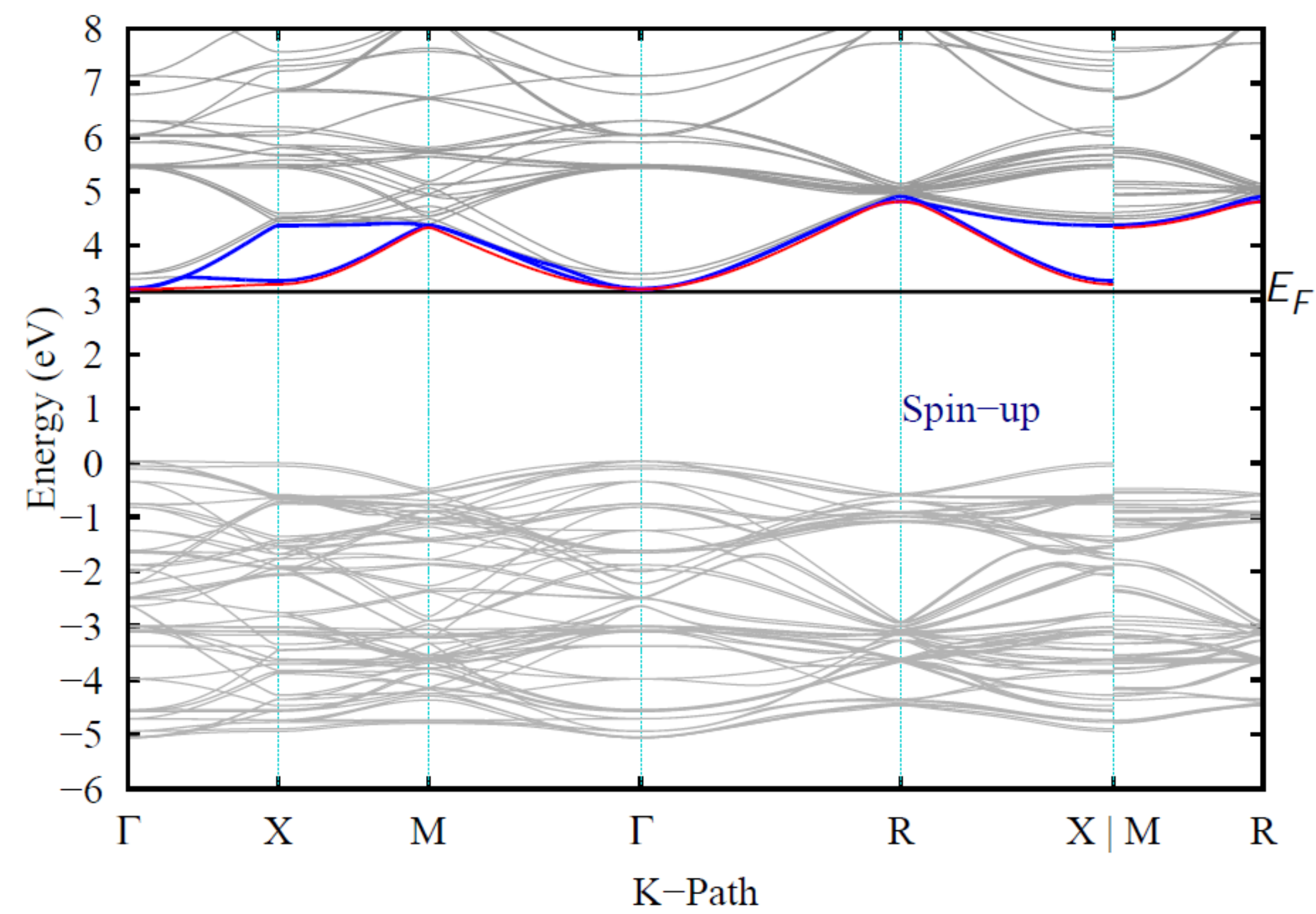} 
\includegraphics[height=0.25\textheight]{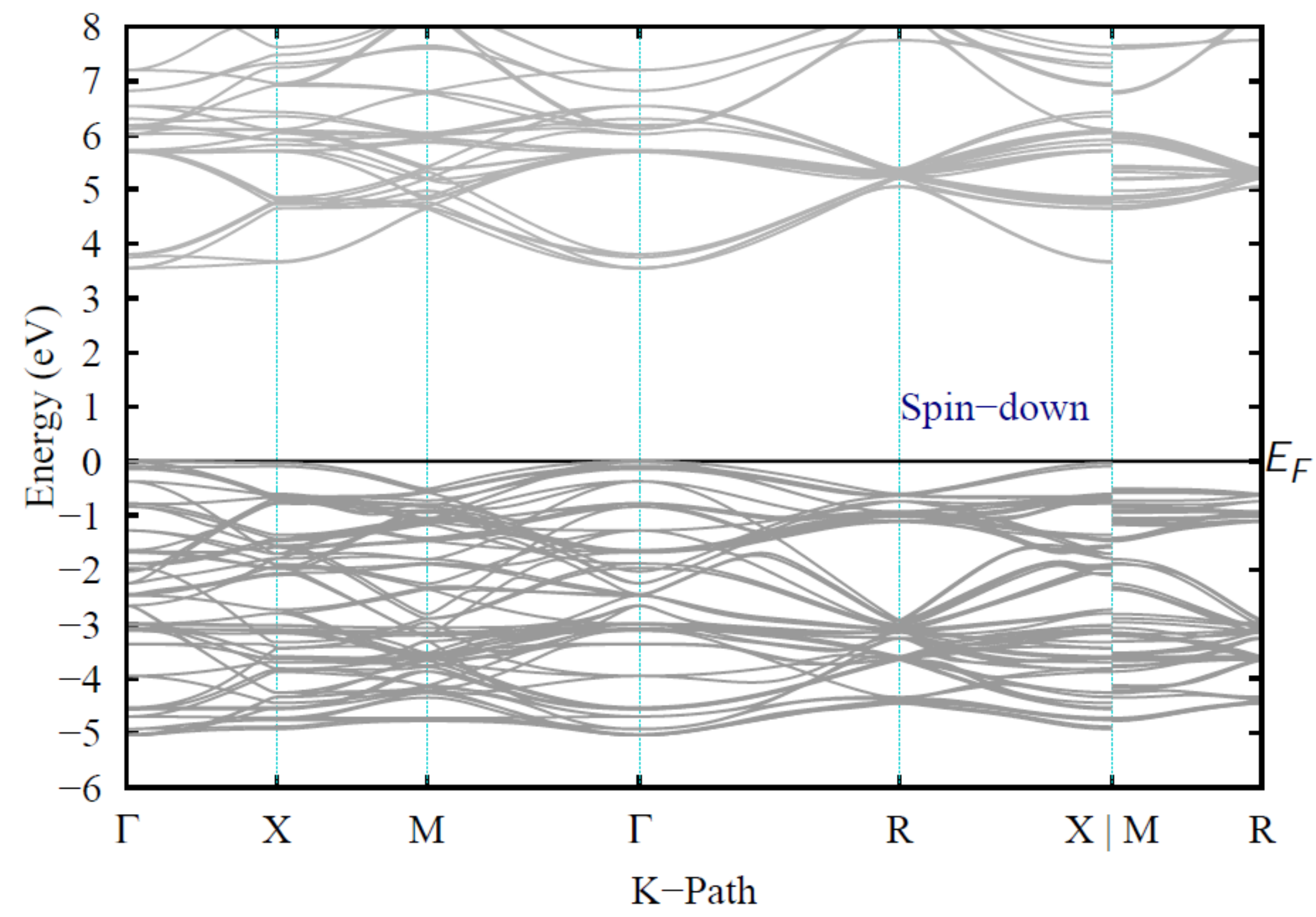}
\caption{\label{fig:La_band}(Color online) Band structure for spin up (Top) and down (Bottom)
electrons in a  2\by2\by2 \STO~supercell doped with La. The Fermi  energy \textit{E$_{F}$} is indicated by a solid black line. For spin up electrons, one of the triply degenerate bands in the  CBM is populated  (shown in red) while  the remaining empty ones are  shown in blue.}
\end{figure}

Figure~\ref{fig:La_band} shows the band structure of
Sr$_{0.875}$La$_{0.125}$TiO$_3$ for spin up and spin down electrons. For the spin
up electrons, the VBM  is almost  triply degenerate with a very small splitting of
7 meV; this is  also the case of the lowest CB, a band which is populated with 
one extra electron and shows a small band broadening (13 meV). The energy
difference between the last VB and the first populated CB is 3.16 eV. The next set
of conduction bands, normally located at 200 meV above the CBM in the bulk STO, is
split into one band a  172 meV above the CBM followed by a doubly degenerate band
100 meV higher. \FE{ The visualization of isosurface of the highest occupied 
 state  at the $\Gamma$ special k-point show that the charge density is not  localized on the defect site, but rather corresponds to the Ti t$_{2g}$ states. The electronic density of states (not shown) also confirms that  CB remains dominated by Ti $3d$ states as it is does in the STO bulk phase;  La $4d$
starts to  contribute to the CBM only at about 2.3 eV above the Fermi  energy. }

For the spin down electrons (bottom of figure~\ref{fig:La_band}) the band structure conserves most of the bulk
band characteristics with triply degenerate CBM and VBM, but the band gap  is
larger than the bulk value, measuring 3.55 eV.  The heavy electron  effective
masses relative to the free electron mass ($m^*_{he}/m_e$) of  the lower band or
heavy electron (he) band  are isotropic and experience a  decrease from the bulk
value of 7.3 to 6.8 in the $\Gamma \rightarrow$X direction, bringing the
$m^*_{he}/m_e$ ratio  closer to the experimentally measured
values~\cite{Ravichandran:2011} of  6-7.1 for 15\% La doping. A smaller decrease
is observed along the other high symmetry directions. 

\section{Summary and conclusions}
\begin{figure}[!h]
\includegraphics[width=0.8\columnwidth]{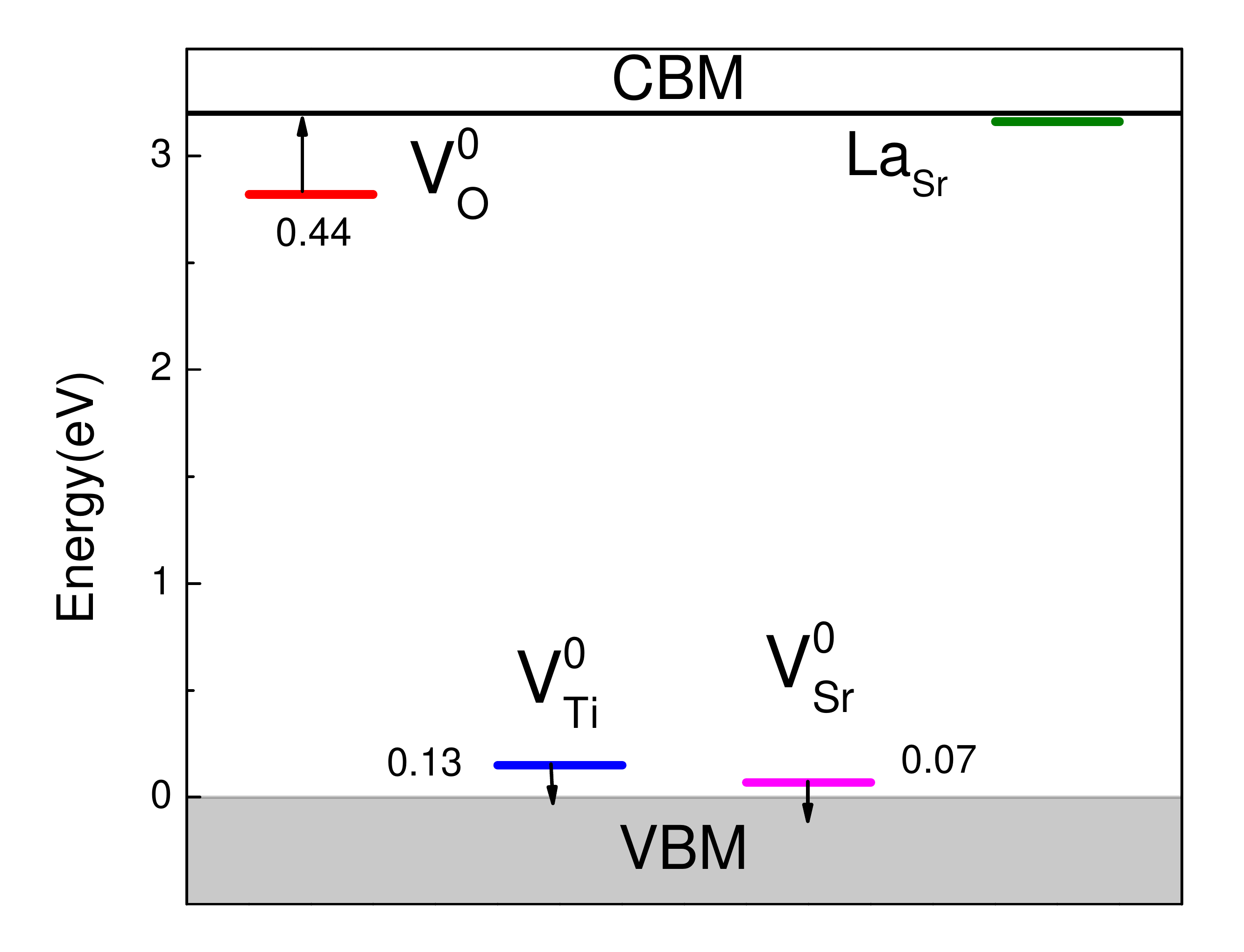}
\caption{\label{fig:defect} (Color online) Schematic representation of  the location of the defect levels in the band gap of \STO~calculated with HSE/SZVP. Numbers and arrows  refer to the location of the defect band  with respect to the nearest bulk-like bands.}
\end{figure}

The structural and electronic   properties of point defects \Vn{O},  \Vn{Ti},
\Vn{Sr} and  La$_{Sr}$ in \STO~have been computed using the HSE screened hybrid
functional. The crystal field splittings of the conduction and valence bands,
resulting from the atomic relaxation around the various defects, were
evaluated. HSE is known to give an accurate description of band gap and VBW of
\STO, which leads to the assumption it will perform also well the defect
energetics, as we have shown here.  In fact, the location of the neutral defect
bands in the band gap of STO~calculated with HSE/SZVP do not suffer from  the
band gap underestimation problems displayed by semilocal functionals, thus
negating the need for further corrections. \FE{The  wavefunction is  localized around \Vn{O} and \Vn{Ti} as shown form the isosurface of  highest occupied orbitals in figure~\ref{fig:iso} and delocalized for the defect states that almost or  completely overlap with host bands like \Vn{Sr} and La$_{Sr}$.} Our calculated defect bands positions
are represented schematically in Figure~\ref{fig:defect}; we present the results
this way for ease of use and comparison to subsequent works. The location of the
defect level in the band gap of STO indicates that \Vn{O}  is probably at the
origin of the blue luminescence of STO, and serves as a double shallow donor
under thermal equilibrium.  This  defect level diagram might serve as a
guideline in the interpretation of photoluminescence
experiments.~\cite{Ravichandran:2011, Liu:2011}  For \STO~doped with \Vn{O}, the calculated conduction band electron effective masses support the proposal of  enhanced mobility along the strain directions.~\cite{Jalan:2011}

Finally, with its two fixed and unmodified parameters, the HSE functional gave
a reliable description of the electronic structure for STO and its defects, in
agreement with the findings in Ref.~\onlinecite{Deak:2010,Deak:2011}. It would
be very interesting to extend the present study for larger supercells followed by the a finite size scaling. This would enable assess the robustness of the present agreement we get with experiment and allow a better comparison by performing double doping with La and \Vn{O}.

\begin{acknowledgments}

This work is supported by the  Qatar National Research Fund  (QNRF) through the National Priorities  Research Program (NPRP  08 - 431 - 1 - 076). We are grateful to the Research computing facilities at Texas A\&M University at Qatar for generous allocations of computer resources.\\

\end{acknowledgments}

\end{document}